\documentstyle[12pt]{article}
\renewcommand{\thesection}{\Roman{section}}

\newcommand{\ovn}{{\pi}}
\newcommand{\pil}[1]{\left\langle #1 \right\rangle}
\newcommand{\lsm}{L$\sigma$M}
\newcommand{\wsub}[1]{_{\mbox{{\scriptsize #1}}}}
\newcommand{\abs}[1]{\left| #1 \right|}

\def\bard{d\kern-2.5pt\raise 3pt\hbox{-}}
 
\title {Nontriviality of the Linear Sigma Model}
\medskip
\author{ L. R. Babukhadia and 
  M. D. Scadron \\ Physics Department, University of Arizona, \\ 
  Tucson, AZ 85721 USA}
 
\date{received 9.11.1998}

\begin{document}
\maketitle

\begin{abstract}

We consider techniques (based on an ultraviolet cutoff) used to prove that 
the pure boson ($\phi^4)_4$ field theory is trivial and apply them instead to 
the dynamically generated quark-level linear sigma model.  This cutoff 
approach leads to the conclusion that the latter field theory is in fact 
nontrivial.


\end{abstract}
\baselineskip 22pt
\section{Introduction}

Owing to the recent observational identification [1] of a nonstrange scalar $\sigma$ meson below 1 GeV, formal field theories discarding such a scalar 
$\sigma$ due to "triviality" theorems (meaning the meson-meson coupling 
$\lambda \to 0$ when cutoff $\Lambda \to \infty$)
should be reanalyzed as well.
In the present paper we show that the quark-level Linear Sigma Model (L$\sigma$M) is  a 
non-trivial field theory in {\em contrast} with the possibly trivial pure boson ($\lambda \phi^4)_4$ 
theory.  Prior studies of $\lambda \phi^4$ field theory using perturbative
and partially nonperturbative methods [2-4] extracted physical constraints
on a scalar meson mass via renormalization group bounds and scaling laws [2].
In a somewhat different manner, there are studies of the triviality problem of
$(\lambda \phi^4)_4$ theory exploiting a new non-perturbative expansion of
the n-point Green's functions [5,6].

Alternatively, one can look at $\lambda \phi^4$ theory also including
fundamental fermions --- the L$\sigma$M.  The key to understanding nontriviality of the quark-level L$\sigma$M is the Goldberger-Treiman
Relation (GTR) which must hold at quark level $m_q = f_\pi g$ to ensure conservation of the axial
vector current.  It turns out that the bosonic sector of the L$\sigma$M theory is totally driven
by the dynamically generated quark mass via the quark-level GTR demanding even fixed and also non-trivial
numerical values for the chiral couplings {\em independent of any UV cutoff} 
[7]. Thus
the presence of the fundamental L$\sigma$M
fermion fields eliminates the possibility for the chiral couplings to vanish and induces a non-trivial
field theory.

In Sec.II we summarize recent results for the dynamically generated quark-level L$\sigma$M.  Then 
in Sec.III we review both perturbative and nonperturbative techniques for solving the problem of triviality for the pure 
boson $(\lambda \phi^4)_4$ theory.  Finally, in Sec.IV we demonstrate nontriviality of the quark-level
L$\sigma$M.  Our results are summarized in Sec.V. 
In the appendix we review regularization schemes for the quark-level L$\sigma$M.

\section{Dynamically generated quark-level L$\sigma$M}

It has been shown recently [7] that the interacting part of the
{\em dynamically generated} SU(2)
quark-level L$\sigma$M lagrangian shifted around the
true vacuum, with expectation values $<\vec{\pi} > = <\sigma > = 0$
is given by:
\begin{equation}
{\cal L}^{Int.}_{L\sigma M} = g^\prime \sigma ( \sigma^2 + \vec  \pi^2) - 
{\lambda \over 4} ( \sigma^2 + \vec\pi^2)^2 + g \bar \psi ( \sigma + i
\gamma_5 \vec \tau \cdot\ \vec \pi ) 
\psi , 
\end{equation}
with the Gell-Mann-L\'evy chiral couplings [8]:
\begin{equation}
g = {m_q \over f_\pi } \ \ \ \ \mbox{and} \ \ \ 
g ^\prime = {m_\sigma^2 \over \rm 2 \it f_\pi } = \lambda f_\pi ,
\end{equation}
and with $m_\pi = 0$.
Here, the chiral-limiting pion decay constant $f_\pi \approx 90$ MeV is 
generated through a {\em logarithmically} divergent quark
loop (fig.1).  Using Feynman rules for fig.1 and the quark-level Goldberger-Treiman Relation
(GTR) $f_\pi g = m_q$, one is led to the following logarithmically divergent gap equation
\begin{equation}
1 = -4iN_c g^2 \int {\bard^4 p \over (p^2-m^2_q)^2} ,
\end{equation}
where $N_c$ is color number and $\bard^4 p \equiv d^4p / (2\pi)^4$.  

A quark mass $m_q$, however, is generated
by the {\em quadratically} divergent tadpole diagram of fig.2.  Once the
\lsm\ is dynamically induced by figs.\ 1 and 2, such divergent
graphs must be
supplemented by $\sigma$ (shifted field) and $\pi$ mediating quark self-energies which sum to zero [7].  Moreover, the resulting L$\sigma$M one-loop
order bubble plus tadpole graphs representing $m^2_\pi$ sum to zero (as they
must by the Goldstone theorem).  In addition, the Lee null tadpole condition
[9], summing the quark plus $\sigma$ plus $\pi$ tadpole graphs to zero
should
also hold in the L$\sigma$M.

  Taking into account dynamically generated meson interactions, one should verify that Lee's null tadpole condition holds for the shifted field $\sigma$. 
This means that the sum of the tadpole graphs of fig.\ 3 must vanish:
\begin{equation}
  \pil{\sigma} = 0 = -8iN_cgm_q \int \frac{\bard^{2l}p}{p^2 - m^2_q} + 3ig' 
  \int \frac{\bard^{2l}p}{p^2 - m^2_\pi} +
  3ig' \int \frac{\bard^{2l}p}{p^2 - m^2_\sigma}.
\end{equation}
In the dimensional regularization approach these three tadpole quadratic
divergences scale respectively like $m^2_q, m^2_{\pi}, m^2_\sigma$ in $2l
= 4$ dimensions.  Then using eqs. (2), one finds that eq.\ (4) requires in
the chiral limit [7] 
\begin{equation}
  N_c (2m_q)^4 = 3m^4_\sigma .
\end{equation}
Moreover, the chiral anomaly (or \lsm ) prediction of 
the $\pi^\circ \to \gamma \gamma$ quark loop 
amplitude $F_{\pi^\circ \gamma \gamma} = \alpha N_c / 3\pi f_{\pi}$, 
leading to a decay rate 
(for $N_c = 3$) of 
$\Gamma_{\ovn ^\circ \gamma \gamma} = m^3_\pi \abs{F_{\ovn ^\circ 
\gamma \gamma}}^2 / 64 \ovn \approx 7.63$ eV, quite close 
to the measured value of 7.74 $\pm$ 0.55 eV, empirically 
fixes $N_c = 3$.  One then sees from eq.\ (5) that the scalar meson mass
\begin{equation}
m_\sigma = 2 m_q
\end{equation}
has been dynamically generated (in agreement with the Nambu-Jona-Lasinio
four-fermion scheme [10]). In fact in ref.[7] the NJL relation (6) in the
context of the L$\sigma$M was obtained using a dimensional regularization 
lemma linking the log-divergent integral in (3) with the 
quadratic-divergent integral in (4) {\em independent of the cutoffs}.  

	Reversing the argument, inputing the NJL relation (6) into 
Lee's null tadpole condition
(5) requires $N_c = 3$.  This circumvents the sometimes-used 
large $N_c$ limit in the discussion of possible triviality of the 
quark-level L$\sigma$M. Even though the three \lsm\ tadpoles of fig.\ 3
and eq.\ (4) sum to zero, the chiral renormalization of the 
massless Goldstone pion is manifested by these tadpoles in a 
different manner.  Specifically, the sum of the quark bubble and 
quark tadpole graphs contributing to $m_\ovn$ vanishes 
because $g' = m^2_\sigma / 2 f_\ovn$ from eq.\ (2) regardless 
of the implied quadratic divergences in (4).  The \lsm\ 
version of the Goldstone theorem is 
then $m^2_\ovn = 0\wsub{qk~loops} + 0\wsub{$\ovn$~loops} 
+ 0\wsub {$\sigma$~loops}~=~0$.  

As for the chiral couplings, the dimensionless meson-quark coupling 
constant $g$ is determined to be 
for $N_c = 3$ [7]
\begin{equation}
g = { 2 \pi \over \sqrt{3} } \approx 3.6276,
\end{equation}
which is compatible with the 
ratio $m_q / f_\pi \approx 320 \ \rm MeV/90 \ \rm MeV  \approx 3.6$ 
arising from the GTR.  Alternatively, making use of the experimental  
couplings $g_{\pi NN} \approx 13.4$ and 
$g_A \approx 1.26$ [1],
we may estimate $g = g_{\pi NN}/3g_A \approx 3.54$, again in a good 
agreement with (7).  Furthermore, the study of the dynamically 
generated quartic 
meson-meson dimensionless coupling $\lambda$ reveals an important
link between $\lambda$ and $g$:
\begin{equation}
\lambda = 2g^2 = {8\pi^2 \over 3} \approx 26.
\end{equation}
The relation $\lambda = 2g^2$ follows from the log-divergent gap equation (3)
which ``shrinks'' the quark-box graph for $\pi \pi$ scattering to the
quartic
$\lambda$-contact interaction in the L$\sigma$M lagrangian [7].
Alternatively, the Gell-Mann-Levy L$\sigma$M relation in (2)
requires $\lambda = m^2_\sigma / 2f^2_\pi$, which reduces to 
$\lambda = 2g^2$ using (6) and the GTR. 
Converting this $\lambda$ to the dimensionless number $8 \pi^2/3$ in 
(8) follows directly from (7).  

We stress that the nonzero numerical values of the meson-quark 
coupling $g$ in (7) and the meson-meson coupling $\lambda$ in (8) 
are obtained in a manner {\em independent} of the implied cutoffs 
in (3) and (4).

It is remarkable that in the 
dynamically generated quark-level L$\sigma$M, the large meson-quark
coupling $g$ and larger meson-meson coupling
$\lambda$ are both completely driven by the fermion sector of the theory via the GTR, 
demanding for them fixed numerical values, (7) and (8) respectively.
Also, L$\sigma$M schemes derived from the chiral symmetry restoration
temperature [11] find that $\lambda \simeq$ 20 (near (8)) at zero
temperature.
It is important to stress that the ``shrinkage'' of quark loops to a point via eq.\ (3) is a $Z = 0$ compositeness condition [7,12]. 
This $Z = 0$ condition merges the \lsm\ field theory when the $\pi$ and 
$\sigma$ are treated as elementary particles with the NJL four-quark-theory 
when the $\pi$ and $\sigma$ are taken as $q\overline{q}$ bound states. 
In 
either theory $m_\ovn = 0$ and $m_\sigma = 2m_q$ in the chiral limit.
Moreover, meson-meson coupling $g'$ or $\lambda$ in eqs.~(2) immediately leads
to the $\sigma$ meson decay width of $\sim 700$ MeV for the mass 
$m_\sigma = 2m_q\sim650$ MeV.

\section{Triviality of $(\lambda\phi^4)_4$ field theory}

There is a strong {\em external} resemblance between the L$\sigma$M and the $(\lambda\phi^4)_4$
quantum field theories.  The recent studies [2-6] (also see refs.~[13])
attempting to prove the triviality of the latter 
theory will motivate us to investigate the question of triviality of the quark-level L$\sigma$M field theory.  First we will briefly summarize results of the $(\lambda\phi^4)_4$ references
[2-6].  The lagrangian of the $(\lambda\phi^4)_4$ field theory in four dimensional space-time is:
\begin{equation}
{\cal L}_{\phi^4} = {1 \over 2} (\partial \phi)^2 + {1 \over 2} \mu^2 \phi^2
- {\lambda \over 4} \phi^4 ,
\end{equation}
with $\mu^2 > 0$ and $\lambda > 0$, corresponding to the spontaneously broken phase.

Also we consider the purely perturbative approach of refs.~[2-4].  Dashen and
Neuberger [2] employed a perturbative (leading log) result for ultraviolet
cutoff $\Lambda$

\begin{equation}
  \frac{1}{\lambda} \gg \frac{3}{2 \pi^2} \ln {\frac {\Lambda}{m_\sigma}},
\end{equation}

\noindent
to obtain an upper bound on the true scalar meson mass $m_\sigma$.  L\"uscher
and Weisz in ref.\ ~[3] calculated the ultraviolet cutoff dependence $\Lambda$
on the renormalized scalar mass and showed that the scaling laws are 
satisfied when

\begin{equation}
  2 m_\sigma < \Lambda < \infty~.
\end{equation}

\noindent
Next, Kimura et al.~[4] reproduced the Dashen-Neuberger relation (10) using the
perturbative renormalization group and also by invoking nonperturbative (but
approximate) Wilsonian renormalization group methods.  In both cases the
bound in eq.~(10) becomes a rough equality.  Then combining eqs.~(10) and (11),
ref.\ [4] deduces that $m_\sigma \le 400$ MeV.  As proposed in refs\ [2-4],
such a scalar mass of order 400 MeV should be considered in an effective
$(\lambda \phi^4)_4$ theory with dimensionless coupling $\lambda$ (in our
eq.~(8)) of order ten.

To return to the question of the triviality limit $\lambda \rightarrow 0$ as
$\Lambda \rightarrow \infty$, we now consider the non-perturbative 
$\delta$-parameter expansion of Bender et al.~in refs.\ [5,6].  Instead of a
conventional perturbative treatment of the Greens functions, they propose
an expansion in a power series of $\delta$ for a $\lambda(\phi^2)^{1+\delta}$
field theory in $d$ dimensions.  The latter theory, as an extension of (9), is
described by the lagrangian:
\begin{equation}
{\cal L}_{\delta} = {1 \over 2} (\partial \phi)^2 + {1 \over 2} \mu^2 \phi^2
- {\lambda \over 4} M^2 \phi^2(\phi^2M^{2-d})^\delta. 
\end{equation}
Here a fixed mass parameter $M$ has been introduced to allow the interaction term to have
the correct dimensions, i.e. to keep $\lambda$ {\em dimensionless} for arbitrary $\delta$
in any space-time dimension.  Obviously, in the limit when $d\to 4$ and $\delta \to 1$, 
eq. (12) reduces to (9) and the parameter $M$ cancels out.  Then one can show
 [5] that the $n$-point
Green's functions can be expanded as a perturbation series in powers of $\delta$:
\begin{equation}
G^{(n)} (x_1,...,x_n;\delta) = \sum^\infty_{k=0} \delta^k g^{(n)}_k (x_1,...,x_n).
\end{equation}

An advantage of this method is twofold.  First, $\delta$ is the {\em only} parameter to be
treated perturbatively and consequently the results obtained from the $\delta$-expansion
(13) are non-perturbative in the {\em physical} parameters of the theory (such as mass and
coupling).  Second, it was demonstrated [6] that when $d \geq 2$, the coefficients of $\delta^{(k)}$
in (13) are less divergent than the terms in the conventional weak-coupling expansion.
However, the $g^{(n)}s$ in (13) still suffer from ultraviolet divergences and thus regularization and 
renormalization of the theory based on (12) and (13) are necessary.

To regularize the divergent expressions for the physical quantities (mass and coupling),
a cut-off $\Lambda$ is introduced in momentum space [6].  The notion of possible
``triviality" of the $(\lambda \phi^4)_4$ field theory generated by the lagrangian (12)
corresponds to a renormalized coupling $\lambda_R \to 0$ as the ultraviolet cutoff
$\Lambda \to \infty$, so that the theory becomes effectively free.

Bender and Jones in ref.\ [6] demonstrated that the triviality of the theory
eq.\ (12) can only follow for 
$d \geq 4$ dimensions, apart from the {\em pathological} case when the 
unperturbed
(scalar) mass $m$ (defined via $m^2 = \mu^2 - {1\over 2} \lambda M^2)$ is greater than the 
cutoff $\Lambda$.
Stated in a reverse manner, one can infer from the Bender-Jones analysis that for a 
{\em nontrivial} $(\lambda \phi^4)_4$ theory, the 
cutoff $\Lambda$ is bounded by the unperturbed scalar mass as
\begin{equation}
\Lambda < m_\sigma.
\end{equation}

The Bender-Jones result for the triviality bound of $(\lambda \phi^4)_4$
theory, namely $m_\sigma < \Lambda$, is a more restrictive conclusion than
eq.~(11) in the sense that it imposes tighter limitations on the scalar mass
that could possibly generate a non-trivial theory.  However, we will show
in what follows that it is the  non-triviality  (pathological) condition 
eq.~(14) (rather
than triviality bound $m_\sigma < \Lambda$) which in fact holds for the 
dynamically generated quark-level L$\sigma$M field theory of Sec.~II.

\section{Linear $\sigma$ Model - a nontrivial theory}

As we saw in Sec.II, the dynamically generated quark-level L$\sigma$M has a special feature~--- 
the scalar mass and the chiral meson couplings are entirely governed by the quark sector of the
theory.  Moreover, the dynamically induced L$\sigma$M is automatically ``chirally renormalized"
[7] due to the L$\sigma$M Gell-Mann-L\'evy chiral couplings (2).  Therefore, the concept of triviality in fact is a 
non-sequitur in the case of the quark-level L$\sigma$M because the dynamically generated
(renormalized) values for the chiral couplings are finite, fixed nonzero numbers much greater than unity:
\begin{equation}
g = {2\pi \over \sqrt{3} } \ \ \ \rm and \ \ \
\lambda = {8\pi^2 \over 3} \, ,
\end{equation}
{\em independent of any UV cutoff}.
These couplings cannot vanish under {\em any} circumstances provided there are fundamental fermion
fields in the theory generating  the GTR and these nontrivial couplings (15).  Consequently, the
quark-level L$\sigma$M is not effectively free, but is instead a dynamically generated nontrivial nonperturbative
field theory.

Nonetheless, one can consider splitting up the quark-level L$\sigma$M lagrangian into ``bosonic"
and ``fermionic" parts to study its ``bosonic" piece alone (in the spirit of refs. [2-6])
but satisfying the NJL scalar mass condition $m_\sigma = 2m_q$.  The results of [2-6]
that were briefly summarized in the preceding section indicate that the question of triviality 
in a pure boson theory, such as a $(\lambda \phi^4)_4$ field theory, crucially depends on the 
relative scales between the ultraviolet (quadratically divergent) 
cutoff $\Lambda$ and the scalar mass $m_\sigma$.
The relations (6)-(8) for the quark level L$\sigma$M were, however, first 
obtained using a dimensional regularization approach. These results are in 
fact independent of both ultraviolet cutoff and regularization.

To achieve consistency with a cutoff approach in the L$\sigma$M, one
needs to evaluate the corresponding divergent integrals with the ultraviolet
cutoff introduced.  We start with the logarithmically divergent gap equation (3) due to fig.1.  Evaluating 
(3) for $N_c = 3$ with a cutoff $\Lambda$ yields:
\begin{equation}
1 = -12 ig^2 \int^{\Lambda} {\bard^4 p \over (p^2-m^2_q)^2} = {3g^2 \over 4 \pi^2}
\left[ \ln \left( {\Lambda^2\over m^2_q} + 1 \right) - {1 \over 1+(\Lambda^2 / m^2_q)^{-1}}
\right] .
\end{equation}
Recalling the numerical value of the quark-meson coupling $g \approx 3.6$ (eq.(7)), one
sees  that eq.\ (16) then suggests $\Lambda^2 / m^2_q \approx 5.3$ or
$\Lambda\approx 750$
MeV for $f_\pi \approx 90$ MeV and $m_q = 2\pi f_\pi/\sqrt{3} \approx 326$ MeV.
This 750 MeV cutoff separates the elementary scalar mass $\sigma$(650)
from the $\overline q q$ bound states $\rho$(770) and $\omega$(783):
$m_{\sigma}<\Lambda$ for this nontrivial L$\sigma$M theory.

Next, we consider the quadratically divergent mass gap equation 
corresponding to fig.2 with
a cutoff $\Lambda$:
\begin{equation}
m_q = {8 iN_c g^2 \over m^2_\sigma}
 \int^{\Lambda} {\bard^4 p\ m_q \over p^2-m^2_q} ,
\end{equation}
in the spirit of the quadratically divergent cutoff approach of NJL [10].
Cancelling out
the constant quark mass $m_q$ and using the NJL or \lsm\  relation $m_\sigma = 2m_q$, eq.(17) implies:
\begin{equation}
1 = i {24 g^2 \over (2m_q)^2}
 \int^{\Lambda} {d^4 p \over (2\pi)^4 } { 1 \over p^2-m^2_q} =
{1 \over 2}\left[ {\Lambda^2\over m^2_q} - \ln \left( {\Lambda^2\over m^2_q} + 1 \right)
\right] .
\end{equation}
This mass gap condition (18) leads to $\Lambda^2/ m^2_q \approx 3.5$ or 
$\Lambda \approx 610$ MeV for $m_q = 326$ MeV. Thus this cutoff
$\Lambda$ in (18) is less than $m_\sigma$(652), whereas $\Lambda$ in (16)
is greater than $m_\sigma$(652).

Although we already know in the L$\sigma$M that $m_\sigma = 2m_q$ as
obtained from either the dynamically generated theory [7] or from the Lee
condition eqs. (4)-(6), we could follow NJL (but in a L$\sigma$M context) 
and simulate the scalar mass
$m_\sigma$ as a $\overline q q$ bound state by computing the quark bubble
and quark tadpole Feynman graphs of figs. 4. Such log-and
quadratic-divergent graphs will be cut off in the ultraviolet region at
$\Lambda$, but this will {\em not} be the 750 MeV cutoff of eq. (16). 
Specifically for $N_f = 2$ one obtains from figs. 4, also using 
$g^\prime = m^{2} _\sigma / 2f_\pi$ and the GTR:
$$
m^2 _\sigma = 16 i N_c g^2 \int ^{\Lambda} \bard^4 p \left[ 
\frac {1} {(p^2  - m ^2 _q )}- \frac {m^2_q} {(p^2-m^2_q){^2}} \right]
\eqno(19a) 
$$

$$
= \frac {N_c g^2\ m^2_q}{\pi^2} \frac {x^2}{1+x}, \eqno(19b)
$$
where $x = \Lambda ^{2} / m^2_q$ is the (four-dimensional) dimensionless
cutoff. Now invoking the meson-quark coupling (7), equations (19) reduce
to
$$
m^2_\sigma = 4m ^2_q  \frac {x^2} {1 + x}, \eqno(20a)
$$
which recovers $m_\sigma = 2 m_q$ if 

$$
x^2 = 1 + x,\;\; \; or \; \; \;\; x = \frac {1+\sqrt{5}} {2} \approx
1.618.
\eqno(20b)
$$

\noindent
Then again using $m_q\approx 326$ MeV, the above cutoff of 
$\Lambda^2 / m^2_q \approx 1.618$ corresponds to $\Lambda\approx 415$ 
MeV.

Note that we have two different cutoffs. The first from the L$\sigma$M
log-divergent gap equation (3) leading to (16) and $\Lambda \approx 750
MeV$ is valid when the $\sigma$(650) meson is treated as an
{\em elementary} {\em particle}. The second cutoff of $\Lambda \approx$
610 MeV or 415 MeV
found from (18) and (20) treats the $\sigma$ meson as a $\overline q q$
{\em bound} {\em state}. Thus it is not surprising that $\Lambda <
m_\sigma$ in the latter cases; it simply means that the $\sigma$ meson can
no longer be treated as elementary when computed via (cutoff) quark loops
from figs.4

Noting that the bosonic part of the quark-level L$\sigma$M is identical with
$(\lambda \phi^4)_4$ field theory provided that $\phi \equiv (\sigma, \vec\pi)$, we can proceed further
and apply the $\lambda \phi^4$ results of ref.\ [5] to the quark-level L$\sigma$M.  Clearly, this L$\sigma$M field theory
falls into the pathological case designated by (14):  the dynamically generated scalar
($\sigma$ meson, in this case) mass $m_\sigma = 2m_q \approx 652$ MeV is 
greater than the cutoff
$\Lambda$ of 415 MeV obtained from (19) and (20), or $\Lambda <m_\sigma$.

\noindent
Therefore, even the ``bosonic" piece of the dynamically
generated L$\sigma$M lagrangian generates a nontrivial field theory in the sense of the Bender-Jones 
condition (14).

It is important to stress the relative inequality structure of eq.\ (14)
and not the implied absolute numerical values in (14) of 415 MeV $<$ 652
MeV.
Specifically, the quark loop calculation of $m_\sigma$ in (19) and (20) 
requires $\Lambda^2 \approx 1.618
m^2_q,$
while the NJL-\lsm\ scalar mass squared is $m^2_\sigma = 4 m^2_q$.  So the
Bender-Jones pathological inequality (14) is always valid regardless of
the size of $m_q$ (i.e.\ 1.618 $<$ 4).  Stated another way, the
Bender-Jones
triviality limit $\Lambda^2 \geq 4 m^2_q$ (as opposed to (14)) can never
be numerically reached from eq.\ (20).

Rather than dealing with the above cutoff approach to triviality
as applied to the (nontrivial) L$\sigma$M, we may instead follow
ref. [7] by starting with the chiral quark model (CQM) {\em massless}
lagrangian
$$
L _{CQM} = \overline{\psi} \lbrack  i\gamma\cdot\partial + g (\sigma + i
\gamma _5 
\vec\tau\cdot\vec\pi)\rbrack \psi + [(\partial \sigma)^2 +
(\partial\vec\pi)^2]/2. \eqno(21)
$$

\noindent
Then the L$\sigma $M lagrangian in (1) is dynamically generated
by subtracting and adding quark and meson mass terms to (21). The former
$-m_q$ and $-m^2_\sigma$ masses then nonperturbatively appear in the 
quark and meson loops of Figs. 2 and 4, while the latter $+m_q$ and 
$+m^2_\sigma$ arise as counterterm masses. More specifically, we
change the sign of $m^2_\sigma$ on the left hand side (lhs) of (19a) and
the
sign of the quadratically divergent first term (17) on the rhs of (19a) as
they
now represent the counterterm $m^2_\sigma, m_q$ masses, respectively. 
However the second term on the rhs
of (19a) must still be computed from the log-divergent gap equation (3).
Then (19a) becomes replaced by 
$$
- m^2_\sigma = - 2m^2_\sigma + 4 m^2_q. \eqno(22)
$$
The unique solution of (22) is the NJL relation $m_\sigma = 2m_q$,
{\em independent} of any ultraviolet cutoff and any regularization
scheme. To recover this NJL result in the cutoff approach of eqs.
(20) requires the Bender-Jones pathological condition 
$\Lambda < m_\sigma$, eq.(14).

\section{Summary}

Thus we must conclude that not only is the dynamically generated quark-level L$\sigma$M quantum field theory
in approximate agreement with data, but also its pure bosonic $\lambda \phi^4$ part is {\em nontrivial}
in the sense that the coupling $\lambda \not \to 0$ as $\Lambda \to \infty$, (indeed $\lambda$ is the 
finite number in (8)).  
In the language of a cutoff theory, the Bender-Jones (pathological)
condition 
[6] $\Lambda < m_\sigma$ appears to be valid for the quark-level
L$\sigma$M when computing $m_\sigma$ in the sense of ref.[6], implying
$\lambda \not \to 0$.  In fact, in a dynamically generated L$\sigma$M, that is dimensionally
regularized [7] with no reference to a cutoff, the bosonic coupling $\lambda$ is $8\pi^2/3$, which
is finite but certainly nonperturbative and nontrivial.

If instead one studied only the bosonic perturbative sector of the L$\sigma$M with the
Gell-\-Mann-\-L\'evy chiral relations (2) requiring $\lambda = m_\sigma^2/
2 f_\pi^2$, one should not expect a small perturbative bound of unity
in (1), i.e. $ |\lambda / 4| < 1$, to place a tight constraint on the $\sigma$
mass.  Rather, in the quark-level L$\sigma$M the meson quartic coupling 
$\lambda$ is quite large $\lambda = 8 \pi^2/3 \sim 26$, and this allows
$m_\sigma$ to be $\sim 650$ MeV when $f_\pi \sim 90$ MeV
(and not $m_\sigma <$ 400 MeV as proposed in refs.\ [2--4]).  One then
might expect such a large contact $\lambda$ coupling to generate a
correspondingly (unphysical) large $\pi \pi$ scattering length, also
incompatible with Weinberg's [14] low energy PCAC analysis.  
Chiral
symmetry, however, requires the $s$, $t$, and $u$ channel $\sigma$ poles 
to cancel off the dominant strength of the large $\lambda \sim 26$ contact 
term, thus recovering the Weinberg $\pi\pi$ scattering behavior [15].

In conclusion then, we suggest that the quark-level L$\sigma$M driven by the
GTR dynamically generates a nontrivial and large nonperturbative meson 
coupling $\lambda \sim 26$ which does not vanish as the cutoff $\Lambda
\rightarrow \infty$~.  This latter field theory should be given serious consideration  
instead of a pure bosonic (and possibly trivial) $\lambda \phi^4$ theory, since a $\sigma$ meson less than 1 GeV now appears in the particle data tables [1].

\medskip
{\bf Acknowledgments}:  The authors are grateful for discussions with 
R.~Delbourgo, H.~F.~Jones, A.~Patrascioiu and for partial support from the
U.~S.~Department of Energy.

\vspace{0.5 in}
\leftline{\Large {\bf Figure Captions}}
\begin{description}
\item{Fig.\ 1.}  Logarithmically divergent quark loop for $f_\pi$.

\item{Fig.\ 2.}  Quadratically divergent graph for $m_q$.

\item{Fig.\ 3.}  Lee sum of \lsm\ tadpole graphs.

\item{Fig.\ 4.}  Nonvanishing contributions to $m^2_\sigma$ in the quark
level L$\sigma$M to one-loop order.

\end{description}

\renewcommand{\thesection}{Appendix: L$\sigma$M Regularization Schemes}
\section{ }
\setcounter{equation}{0}
%%%\newcounter{blah}
%%%\setcounter{blah}{1}
%%%\renewcommand{\theequation}{\Alph{blah}.\arabic{equation}}
\renewcommand{\theequation}{A.\arabic{equation}}

In order to convince the reader that the quark-level L$\sigma$M is completely
free of any (both logarithmic and quadratic) singularities, we review 
refs.~[7] for dimensional and Pauli-Villars regularization schemes.
For dimensional regularization in $2l$ dimensions one expresses the log- and
quadratic-divergent integrals as:
\begin{equation}
\int {\bard^{2l} p / (p^2-m^2_q)^2} = 
  i\Gamma(2-l)(m_q^2)^{l-2} / (4\pi)^l,
\end{equation}
\begin{equation}
\int {\bard^{2l} p / (p^2-m^2_q)} = 
  -i\Gamma(1-l)(m_q^2)^{l-1} / (4\pi)^l.
\end{equation}
Then in the four-dimensional limit ($l\rightarrow2$), the difference
between these two divergent integrals is in fact \emph{finite}: 
\begin{equation}
\int \bard^4 p \left[ \frac{m_q^2}{(p^2-m^2_q)^2} - 
		     \frac{1}{p^2-m^2_q} \right] =
	\lim_{l\rightarrow2} \frac{im^{2l-2}_q}{(4\pi)^2}
	\left[\Gamma(2-l)+\Gamma(1-l)\right] = - \frac{im^2_q}{(4\pi)^2},
\end{equation}
because of the mathematical \emph{identity} 
$\Gamma(2-l)+\Gamma(1-l)\rightarrow-1$ as $l\rightarrow2$. 
This dim. reg. lemma follows from the gamma function property
$z\Gamma(z)=\Gamma(z+1)$.

The above dim. reg. regularization (A.3) also holds for analytic, zeta
function and Pauli-Villars regularization schemes [7]. Specifically, for 
Pauli-Villars regularization, one expresses the difference of the log-
and quadratic-divergent integrals in (A.3) as
\begin{equation}
\int \bard^4 p \left[ \frac{m_q^2}{(p^2-m^2_q)^2} - 
		     \frac{1}{p^2-m^2_q} \right] =
\int \frac{\bard^4p}{p^2} \left[ -1 + \frac{m^4_q}{(p^2-m^2_q)^2} \right].
\end{equation}
The identity (A.4) can be verified by partial fractions of the integrands before
the infinite integrals in (A.4) are performed. For Pauli-Villars regularization, 
introduce an ultraviolet cutoff $\Lambda$ on the right-hand-side of (A.4) and
sum over auxiliary massive fermions (masses $M_j$) with probabilities $c_j$.
Then (A.4) becomes:
\begin{equation}
\int \bard^4 p \left[ \frac{m_q^2}{(p^2-m^2_q)^2} - 
		     \frac{1}{p^2-m^2_q} \right] =
\sum_{j} ic_{j}(\Lambda^2 - M_{j}^{2})/(4\pi)^2.
\end{equation}
Applying the Pauli-Villars sum rules [16] $\sum c_j = 0$, 
$\sum c_{j} M_j^{2} = m^{2}_q$, eq.~(A.5) reduces to $-im^2_{q}/(4\pi)^2$.
Thus (A.5) becomes \emph{precisely} the dim. reg. lemma (A.3), only now
found from the Paulli-Villars regularization scheme.

The fact that the difference between the quadratic and the logarithmic
divergences is a finite number \emph{independent} of any particular 
ultraviolet cutoff $\Lambda$, leads to the cutoff-independent results 
$m_{\sigma}=2m_q$ and $g=2\pi/\sqrt{3}$. The latter also implies that
the quartic meson coupling $\lambda$ ($\lambda=2g^2$) has a finite 
non-zero value which is \emph{independent} of the ultraviolet cutoff.

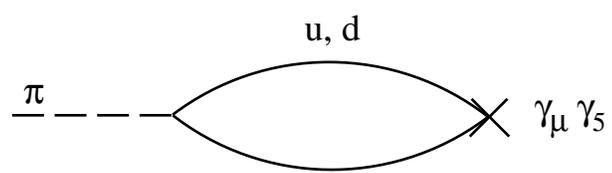
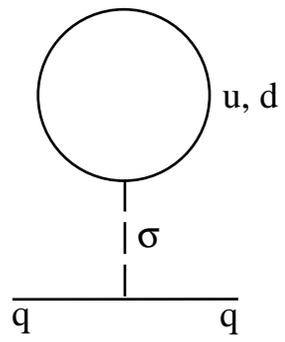
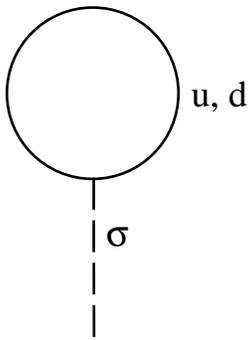
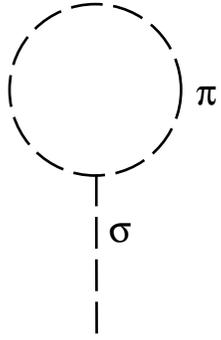
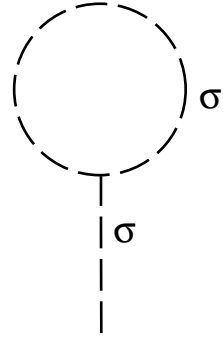
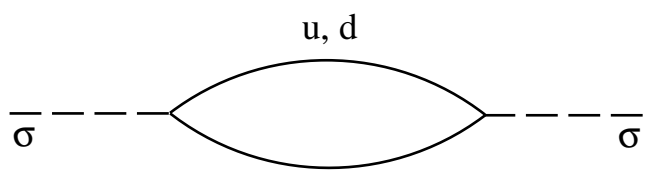
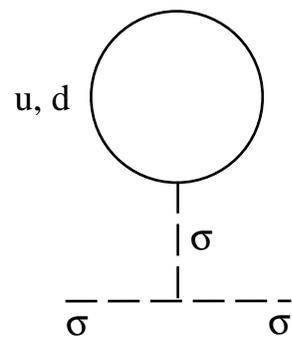
\end{document}